%Paper: astro-ph/9305002
%From: BEST@sns.ias.edu
%Date: 06 May 1993 09:27:28 -0400 (EDT)

\magnification=\magstep1
\baselineskip 8truemm plus 1truemm

\def\lap{\lower.5ex\hbox{$\; \buildrel < \over \sim \;$}}
\def\gap{\lower.5ex\hbox{$\; \buildrel > \over \sim \;$}}

\tolerance=10000

\centerline{\bf Visibility of Gravitational Lenses and}
\centerline{\bf the Cosmological Constant Problem}
\bigskip
\centerline{M. Fukugita$^{(1)}$\footnote*{On
leave from Yukawa Institute, Kyoto
University, Kyoto 606, Japan} and P. J. E. Peebles$^{(2)}$}
\centerline{Institute for Advanced Study$^{(1)}$, Princeton, NJ 08540}
\centerline{Department of Physics, Princeton University$^{(2)}$,
Princeton, NJ 08544}
\bigskip

\centerline{\bf Abstract}
\medskip
Recent observations suggest appreciable star formation activity in
early-type galaxies down to redshift $z\sim 0.5$. If so, there is
likely to be dust in these galaxies.  We consider the possibility that
obscuration by dust can reconcile the observed frequency of
gravitational lensing of quasar images with the considerably
larger rate predicted in a low density cosmologically flat
universe dominated by a cosmological constant.
\medskip

\centerline{\bf I. Introduction}

The analysis
of the Hubble Space
Telescope (HST) Snapshot Survey for
the gravitational lensing of quasars by foreground
galaxies (Bahcall et al., 1992a, b;
Maoz et al. 1993; Maoz \&\  Rix 1993, hereafter MR) confirms the
indications from previous discussions
that the frequency of gravitational lensing favours the Einstein-de
Sitter universe (density parameter
$\Omega =1$ and cosmological constant $\Lambda =0$) rather than a
flat universe dominated by the cosmological constant
(Fukugita, Futamase \&\
Kasai 1990; Turner 1990; Fukugita \&\ Turner 1991, hereafter FT;
Fukugita {\it et al.} 1992, hereafter FFKT; and Kochanek
1992).  If the universe were Einstein-de~Sitter, the frequency
of gravitational lensing at separations
$\theta\sim 1$ arcsec is about what would be expected from the known
properties of massive early-type galaxies, while the lensing
rate at $\theta\sim 2$~arcsec is greater than would be predicted
under the assumption that the mass in a typical bright elliptical
galaxy is distributed like the starlight. The
natural way to remove this discrepancy assumes
bright ellipticals typically have massive dark halos with velocity
dispersion significantly larger than that of the central stars.

If the universe were cosmologically flat with mass density
$\Omega =0.1$ times the Einstein-de Sitter value, on the other hand, the
predicted lensing frequency at
$\theta\sim 2$~arcsec would agree with  the observations, within
the considerable uncertainty in the abundance of galaxies with
the large velocity dispersions which produce well-separated
images, but the predicted lensing rate at
$\theta\sim 1$~arcsec certainly would be much
too large. Ratra \&\ Quillen (1992) have pointed out that, for fixed
$\Omega$ in a cosmologically flat universe, the predicted lensing
rate is reduced by allowing time evolution of an effective
cosmological ``constant'' represented by the energy density
of an evolving dissipationless cosmic field. In this paper we consider a
possible astronomical explanation for the relatively low observed
lensing rate, that dust in the young galaxies at
redshifts $z\gap 0.5$ obscured many lensing events.

The Einstein-de~Sitter cosmology is of particular interest because
it agrees with Dicke's (1970) coincidence argument, that if the
universe is not Einstein-de~Sitter then it is curious
that we have appeared just at the epoch of transition away from
matter-dominated expansion. A universe in which space
curvature and a cosmological constant both are important at the
present epoch requires
a double Dicke coincidence, which seems quite unlikely. Thus the
interesting low density models have
neglible space curvature or else negligible
$\Lambda$. The former is preferred because (1) it agrees with the
requirement of inflation as an explanation for the large-scale
homogeneity of the observed universe (Guth 1981), (2) it allows
the expansion time to exceed the Hubble time $H_o{}^{-1}$, and
(3) in this model the growth of linear density perturbations is
not much suppressed  relative to the
Einstein-de~Sitter case (Peebles 1984). The second point would be
of particular importance if Hubble's constant proved to be close
to the larger of the currently discussed values,
$H_0=80$~km s$^{-1}$~Mpc$^{-1}$ (Jacoby et al. 1992;
Fukugita, Hogan \& Peebles 1993). However, at the bound found by
MR from the HST lensing survey, $\Omega =0.3$, the
cosmologically flat model gives
$H_ot_o=0.96$ (compared to $H_ot_o=0.81$ at $\Lambda =0$ and the
same $\Omega$), which would be of little help for the cosmic age
problem. Thus it would be of considerable interest to know
whether the lensing rate can be reconciled with a
cosmologically flat model, with a constant $\Lambda$ and
$\Omega = 0.1$.

In the next section we review the indications of significant
star formation in galaxies at $z\approx 0.5$. If stars are
forming, one suspects there is dust in the gas out
of which the stars formed and in the gas expelled by massive
stars. We discuss the possible effect of the dust
on the observed lensing rate under two crude models. The
first simply assumes dust made all galaxies opaque at
$z>0.5$. The second assumes that for $z>0.5$
the gas-to-star ratio in young bright
ellipticals is that
characteristic of present-day spirals, and the dust-to-gas
ratio is that of the Milky Way. In either of the two cases
early-type galaxies at $z<0.5$ are assumed to be transparent.
Table 1 shows the effect of these dust models on
the lensing rate in cosmologically flat cosmologies.

\medskip

\centerline{\bf II. Dust in Young Galaxies?}\par
\nobreak
We are not aware of any previous discussion of the possible
effect of dust on the observed rate of gravitational lensing.
The lack of serious consideration of this issue is quite
reasonable, because arcsecond gravitational lensing of quasar
images is thought to be dominated by massive early-type galaxies,
and few such nearby galaxies show evidence of dust
(Centaurus A being the prominent exception). Furthermore,
we do not even have a clear picture of the typical obscuration in
present-day spiral galaxies (Burstein, Haynes \&\ Faber 1991).
Pei, Fall \&\ Bechtold (1991) find evidence of obscuration by
dust in damped Lyman-$\alpha$ clouds, but the redshifts are well
above those of observed gravitational lenses, and the nature of
these young galaxies, whether protospirals or protoellipticals or
dwarfs, is not clear. Thus we have little empirical evidence on
which to base a model for the dust in young ellipticals. As
emphasized by Heisler \&\ Ostriker (1988) and in earlier
references therein, however, there are good astronomical reasons
to think young galaxies are dusty.

The most familiar observation suggesting there is gas in galaxies
at redshifts $z\sim 0.5$ is the Butcher-Oemler effect (Butcher
\&\ Oemler 1978, 1984), that the fraction of cluster
members with the colors of spiral galaxies is considerably larger
at $z\sim 0.35$ to 0.5 than at the present epoch.
A similar trend is observed in the
field (Koo \& Kron 1992).
The idea that the colors are due to increased star formation activity
is corroborated by the observation that at $z\approx 0.3-0.4$
an appreciable fraction of cluster galaxies exhibit the Balmer
absorption lines characteristic of A stars which
would have formed at redshifts not much earlier than the epoch of
observation  (Gunn \&\ Dressler 1988; Dressler \& Gunn 1992).
The HST observations by
Dressler, Oemler, Gunn and Butcher (Dressler 1993)
of a cluster at $z\sim 0.4$
suggest the Butcher-Oemler effect in this system arises from an
unusually large
abundance of spiral cluster members. The spirals may
since have faded, or possibly this is an unusual
cluster. Since the galaxy images in this cluster tend to be
irregular, however, suggesting the merging rate is large,
one might expect there is a large rate of occurrence
of the Centaurus A phenomenon resulting from mergers of spirals
in the bright ellipticals.

Perhaps more direct evidence of gas in young giant ellipticals is the
alignment of the radio and optical images. This
effect appears at redshift $z\sim 0.7$, and most interpretations
involve gas either in star formation or in scattering of starlight
along the radio jet (Daly 1992; McCarthy 1993).

We shall estimate the possible effect of dust on the observed
lensing rate under the
following assumptions. (a) We follow Silk (1993), who notes that
the star formation accommodated by a gas component
which is about 2~percent of the stellar mass would account for the
general evolution of galaxy colors, and the incidence of A-star
spectra, at $z\gap 0.5$. (b) We assume the conversion from gas to
dust to extinction is the same as in the Milky Way. (c) We
approximate the space distribution of the
gas by a truncated King model.

\medskip
\centerline{\bf III. Gravitational Lensing Frequencies}\par
\nobreak
For the analysis of gravitational lensing, galaxies are usefully
approximated by singular isothermal spheres. In this model the
lensing frequency, neglecting obscuration, is (Turner, Ostriker \&
Gott 1984; FT)
$$\tau=\int^{z_s}_0 F (1+z_L)^3 \bigg({D_{OL}D_{LS} \over R_0 D_{OS}}
\bigg)^2 \bigg({1 \over R_0}{dt \over dz_L}\bigg) dz_L~, \eqno{(1)}$$
where $D_{OL}$, $D_{OS}$, $D_{LS}$ are the angular size distances
among the observer $O$, lens $L$ and source $S$, $R_0=1/H_0$ is the
Hubble distance, the quasar and lens are at redshifts $z_S$ and
$z_L$, and the parameter $F=16\pi^3n_0\sigma^4R_0^3$ depends on
the comoving number density $n_0$ of galaxies and the line of
sight velocity dispersion $\sigma$ in the isothermal gas sphere
model for the mass distribution.
FT find $F_E=0.019\pm 0.008$ for ellipticals,
$F_{S0}=0.021\pm 0.009$ for S0s and $F_S=0.007\pm 0.002$
for spirals,
where dark halos are assumed to give velocity dispersion $(3/2)^{1/2}$
that of stars.
With these numbers in equation (1), the lensing
probability for the 502 quasars in the HST sample (Maoz et al. 1993)
is $\tau=1.06\times 10^{-3}$ per quasar
for the Einstein-de~Sitter model
and $7.04\times 10^{-3}$ for a $\Lambda$-dominated
flat universe with $\Omega_0=0.1$. The dominant
contributions come from E+S0 galaxies, in particular those with redshifts
$\sim (1/3)z_S$ for the Einstein-de~Sitter model and $\sim
(1/2)z_S$ for the low density case. The relatively large
predicted frequencies are due to a large average redshift of the
HST quasar sample, $\langle z_S \rangle = 2.2$.

The value of $\tau$ from equation (1) must be corrected for (i)
finite core size  ($\sim 0.63\times$), (ii) angular selection
effect ($\sim 0.95\times$) for
E and S0,  and (iii) the magnification bias ($\sim 9.1\times$).
With these corrections, and neglecting obscuration, the predicted
lensing rate for the 502 quasars is shown
in the second row of Table 1, and can be compared to the observed
rate in the first line.
We list separately the lensing events at
angular separation $\theta<1$~arcsec, which are insensitive
to the presence or absence of a dark halo, and
the events at $\theta >2$~arcsec which dominantly arise from
a dark halo with a large $\sigma$.

The predictions in the second row of the table, which are based on the
procedures detailed in FFKT, are generally similar to those of MR, except
that the MR prediction for the lensing rate at $\theta > 2$ arcsec
is a factor of two lower. This arises from the
differences in the models for the dark halo, and the spread
may be a useful measure
of the uncertainty in the predictions at large $\theta$. We
conclude that, within this uncertainty, the lensing rate at
$\theta >2$~arcsec is not inconsistent with the low density model.
The issue to be considered next is the possible effect of
obscuration at the smaller impact parameters for lensing events
at $\theta\sim 1$ arc sec, where the observed lensing rate
definitely is
lower than would be expected in the low density model.

\medskip
\centerline{\bf IV. Models for Obscuration}\par
\nobreak
The third line in the table shows the effect of
obscuration in an extreme case, in which galaxies are transparent at
$z<0.5$ while at larger redshifts dust completely obscures all
events by blocking at least one of the lines of sight. We
represent this by placing a cutoff in
the integral in equation (1) at $z=0.5$. The
reduction factor for the low density case is relatively
large because the distribution of lens redshifts peaks around
$(1/2)z_S$, substantially deeper than in the Einstein-de~Sitter
model. Also, the redshift cutoff doubles the mean ratio
$\langle D_{LS}/D_{OS}\rangle$, and correspondingly increases the
typical image separation. Thus in the low density model the peak of
the probability distribution $dP/d\theta$ is $\theta\sim 2^{\prime\prime}$
with the cutoff, compared to $\theta\sim 0.9^{\prime\prime}$ in
the completely unobscured case.

A possibly  more realistic approach uses the ratio
of dust to starlight observed in nearby objects. We again assume the
galaxies responsible for lensing are transparent at $z<0.5$, and
at larger redshifts we assign each young early-type galaxy a fixed
gas mass with density as a function of radius
$$
        \rho (r) = \rho _o[1+(r/r_c)^2]^{-3/2}\ ,\eqno(2)
$$
at $r<10$~kpc, and $\rho (r)=0$ at larger radii.
To normalize this expression we assume
the HI gas mass is 2\%\ of the total stellar mass
based on the nominal mass-to-light
ratio $M_{\rm stellar}/L=5$ and the galaxy luminosity calibrated
to $M^*_{B_T^0}=-20.4$ at $H_0=80$~km~s$^{-1}$~Mpc$^{-1}$. We
use the linear relation between the HI column density and
obscuration in the B-band in the Milky Way (Burstein \& Heiles
1978),
$$
        A_B= 4E_{B_V}= 0.79 N_{21}-0.22\ ,\qquad
        N_{\rm HI}=10^{21}N_{21}\hbox{ cm}^{-2}\ .
                \eqno(3)
$$
The computed effect on the observed lensing probability depends
only weakly on the choice of the core radius in equation (2) in
the range $r_c=0.5$ to 5~kpc.

We consider two aspects of the effect of obscuration.
Let us confine ourselves to the case of a low density universe.
First, in lensing events with large magnification
the impact parameters are
at $r_p\sim 4$~kpc for $\theta =1''$ and
$z_L=0.5$, and the quantity of interest is the typical value
of the extinction $A$ at this distance from  the center of a
young bright elliptical galaxy. Second, if the impact
parameters of the two images are substantially
different the image with the smaller
impact parameter is less strongly magnified and would be expected
to be more strongly obscured. Obscuration thus tends to push the
magnitude difference of the images beyond the threshold for
detection.

The effect of the mean obscuration $\langle A\rangle$ can be
included in the expression for the factor $B(m)$ by which
the probability for observing lensing of a faint background
quasar event is enhanced by its magnification:
$$B(m)={\int^\infty_{\Delta_0} N_Q(m+\Delta-\langle A\rangle)P(\Delta)
    d\Delta
    \over    N_Q(m)}~.\eqno{(4)}$$
Here $N_Q(m)$ is the differential quasar number-apparent
magnitude luminosity function, and
$P(\Delta)=7.37\times 10^{-0.8\Delta}$ (for $\Delta\ge
\Delta_0=2.5 \log 2$) (FT),

With the above model and numbers we find that the mean extinction
is $A=0.8$, and that this reduces the effective magnification
enhancement factor $B(m)$ in the HST sample by 60\%.

Next, let us consider the magnitude difference of
the images. In the singular isothermal sphere model the two images
of a quasar at true angular distance $\theta_S$ from the
centre of a galaxy appear at angles $\theta_1 =\beta+\theta_S$,
$\theta_2 =-\beta+\theta_S$
where $\beta=4 \pi \sigma^2 D_{LS}/D_{OS}>\theta _s$
for two images.
The magnitude difference resulting from the difference in
magnification of the two images is
$\Delta m=2.5 \log [(\beta +\theta _S)/(\beta -\theta _S)]$
with the image at $\theta_2$ fainter.
In the HST survey the detection limit for two images
depends on the image separation angle.
For the typical separation angle 1.7~arcsec for $L^*$ galaxies
(for $z_S\sim 2$ and $z_L\sim 1$) the detection limit is
$\Delta m\le 3.5$~mag (Bahcall et al. 1992a).
This means the survey would miss a lensing
event if the difference $\Delta A(\theta _S)$ of obscuration of
the images at the smaller and larger impact parameters satisfied
$$
        \Delta m(\theta _S) + \Delta A(\theta _S) \gap 3.5\ .
                        \eqno(5)
$$
We find that in our dust model obscuration in an $L_\ast$
galaxy makes the magnitude difference larger than this
detection limit at $\theta _S\lap (0.5\hbox{ to }0.6)\beta$,
leaving a survival
probability $\sim 0.3$ for observed lens events.

The expected lensing frequency is the sum of the unobscured
rate, $\tau(0;0.5)$, for low redshift lenses, and the obscured
rate for larger redshift, $0.4\times 0.3\times \tau(0.5;z_s)$,
the first factor being the suppression of the
magnification bias (eq. [4]), the second the effect of the
difference of extinction of the images (eq. [5]), and the last
the unobscured rate. The results are presented in the fourth entry
of Table 1, along with the results of a similar calculation
for the Einstein-de Sitter universe.

\medskip
\centerline{\bf V. Discussion}\par
\nobreak
The gravitational lensing rate is a remarkably sensitive
test of cosmological models, but, in common with the other tests,
it does require an understanding of the astronomy of the
objects. The HST results are in line with
earlier samples in indicating that the distribution of angular
separations $\theta$ in lensing events is broader than might have
been expected in an Einstein-de~Sitter universe. This is not a
serious problem, of course, because it is easy to imagine
lensing events at large $\theta$ has been enhanced by
massive dark halos while the lensing frequency
at small $\theta$ have been slightly obscured by dust.
An open cosmological model with negligible
$\Lambda$ and $\Omega =0.1$ predicts lensing rates only modestly
greater than in the Einstein-de~Sitter model. This would seem to
be easy to accommodate by adjusting the parameters in the
same manner. The $\Omega =0.1$
cosmologically flat model requires a
considerably larger adjustment. We have attempted to show,
however, that the adjustment is not unreasonable in view of
the expected properties of the active young galaxies
observed at $z\gap 0.5$.

Observations in the infrared may test the hypotheses that
optical lensing events have been obscured by dust.
Since dust may be ignored in radio-selected gravitational lenses,
the radio surveys may offer a key test (Turner 1993). For example,
the optical images of the lensed quasar in MG0414+0534
are unusually red ($g-r\approx 2.3$), and the color varies
from image to image ($r-z\approx 2.4$ to 1), suggesting obscuration
as much as 5 mag in the B-band, if the unusual colors are to be
ascribed to dust in the lensing galaxy (Hewitt et al. 1992). It may
be significant also that radio-selected lenses generally are
rather faint in the optical band, compared with optically
selected lenses, as would be expected if there were appreciable
dimming by dust. The preliminary report from the VLA lens
survey (Hewitt et al. 1989) presents four lenses out of 4000
sources. The rate is comparable to that of optically
selected lenses (2 to 9 out of 4300 in the Hewitt \& Burbidge 1987, 1989
catalog).  However, redshifts of quasars in radio surveys are
known only for a part of the sample,
and the average redshifts are not as high as those in the optical
sample, which makes the
interpretation more sensitive to astronomy and less sensitive to
cosmology.  Since the
identification of radio lensing events is complicated by the
intrinsic structures of the objects, it will
not be easy to identify the abundance of objects capable of being
imaged as double images.  One must look for objects with more
complicated geometry, or
the characteristic radio ``Einstein rings''. The
progress in the radio surveys for these events
certainly will be followed with great interest.

\bigskip
\bigskip
We are grateful to C. J. Hogan, D. Maoz, H.-W. Rix, and J. Silk
for useful discussions, and to J. Hewitt and
E. L. Turner for their instruction on the radio detection of
gravitational lensing. This work was supported in part at
Princeton University by the US National Science Foundation.

\vfill\eject

\def\ref{\hangindent=5ex\hangafter=1}
\parskip=0pt
\parindent=0pt
\noindent{\bf References}

\ref Bahcall, J. N. et al. 1992a, ApJ 387, 56

\ref Bahcall, J. N. et al. 1992b, ApJ 392, L1

\ref Burstein, D., Haynes, M. P. \&\ Faber, S. M. 1991, Nature
        353, 515

\ref Burstein, D. \&\ Heiles C. 1978, ApJ 225, 40

\ref Butcher, H. \&\ Oemler, A. Jr. 1978, ApJ 219, 18
        and 226, 559

\ref Butcher, H. \&\ Oemler, A. Jr. 1984, ApJ 285, 426

\ref Daly, R. A. 1992, ApJ 386, L9

\ref Dicke, R. H. 1970, {\it Gravitation and the Universe}
(Philadelphia:
        American Philosophical Society)

\ref Dressler, A. 1993, in Proceedings of the Milan Conference
         on Observational Cosmology, 1992 (to be published)

\ref Dressler, A. \&\ Gunn, J. E. 1992, ApJS 78, 1

\ref Fukugita, M., Futamase, T. \&\ Kasai, M. 1990, MNRAS, 246,
        24P

\ref Fukugita, M., Futamase, T., Kasai, M., \&\ Turner, E. L. 1992,
        ApJ 393, 3 (FFKT)

\ref Fukugita, M., Hogan, C. J. \&\ Peebles, P. J. E. 1993,
        submitted to Nature.

\ref Fukugita, M. \&\ Turner, E. L. 1991, MNRAS 253, 99 (FT)

\ref Gunn, J. E. \&\ Dressler, A. 1988, in Towards
        Understanding Galaxies at Large Redshifts,
        eds. R. G. Kron and A. Renzini (Dordrecht: Kluwer), 227

\ref Guth, A. 1981, Phys Rev D 23, 347

\ref Heisler, J. \&\ Ostriker, J. P. 1988, ApJ 332, 543

\ref Hewitt, A. \&\ Burbidge, G. 1987, ApJS 63, 1

\ref Hewitt, A. \&\ Burbidge, G. 1989, ApJS 69, 1

\ref Hewitt, J. N. et al. 1989, in Gravitational Lenses, Lecture
   Note in Physics, 330, eds. J. M. Moran et al.
   (Springer-Verlag, Berlin), 147

\ref Hewitt, J. N., Turner, E. L., Lawrence, C. R., Schneider, D. P.,
    \&\ Brody, J. P. 1992,  AJ 104, 968

\ref Jacoby, G. H. et al. 1992, Publ ASP, 104, 599

\ref Kochanek, C. S. 1992, ApJ 394, 1

\ref Koo, D and Kron, R. G. 1992, ARA\&A 30, 613

\ref Maoz. D. \&\ Rix, H.-W. 1993, ApJ, in press (MR)

\ref Maoz, D. et al. 1993, ApJ, 402. 69

\ref McCarthy, P. 1993, ARA\& A 31, in press

\ref Peebles, P. J. E. 1984, ApJ 284, 439.

\ref Pei, Y. C., Fall, S. M. \&\ Bechtold, J. 1991, ApJ 378, 6

\ref Ratra, B. \&\ Quillen, A. 1992, MNRAS, 259, 738

\ref Silk, J. 1993, private communication

\ref Turner, E. L. 1990, ApJ 365, L43.

\ref Turner, E. L. 1993, private communication

\ref Turner, E. L., Ostriker, J. P. \& Gott, J. R. 1984, ApJ 284, 1

\vfill\eject

\def\hrulefill{\leaders\hrule height0.6pt\hfill\quad}
\centerline{TABLE 1}
\smallskip
\centerline{Models for the HST Lensing Statistics}
\bigskip
\hbox to \hsize{\hss
\vbox{
\hrule height 0.6pt
\vskip 2pt
\hrule height 0.6pt
\halign{\strut
        #\hfil & \hfil\ #\ \hfil & \hfil\ #\ \hfil & \hfil\ #\ \hfil &
        \hfil\ #\ \hfil & \hfil\ #\ \hfil & \hfil\ #\ \hfil \cr
\noalign{\vskip 6pt}
 & \multispan{3}\hfil $\Omega = 1,\ \lambda = 0$\hfil
        & \multispan{3}\hfil $\Omega = 0.1,\ \lambda =0.9$\hfil\cr
\noalign{\vskip -6pt}
 & \multispan{3}\quad\hrulefill & \multispan{3}\quad\hrulefill \cr
 & $N(<1 '')$  & $N(1-2'')$ & $N(>2 '')$  & $N(<1 '')$
 & $N(1-2'')$ & $N(>2 '')$ \cr
\noalign{\vskip 6pt}
\noalign{\hrule height 0.6pt}
\noalign{\vskip 6pt}
Observed & 1 & 1 & 2 & 1 & 1 & 2 \cr
No Dust  & 0.9 & 0.9 & 0.9 & 5.8 & 5.9 & 6.1 \cr
No events at $z>0.5$ & 0.2 & 0.3 & 0.7 & 0.3 & 0.6 & 1.6 \cr
Dust model & 0.3 & 0.4 & 0.7 & 0.9 & 1.1 & 2.1 \cr
}
\vskip 3pt
\hrule height 0.6pt
\vskip 3pt
}
\hss}

\end